\documentclass[final,3p,times]{elsarticle}
\usepackage{amssymb}
\usepackage[fleqn]{amsmath}
\usepackage{amsthm}
\usepackage{subfigure}
\usepackage{graphicx}

\numberwithin{equation}{section}
\biboptions{numbers,sort&compress}
\usepackage{hyperref}

\journal{}

\begin{document}

\begin{frontmatter}

\title{A new integrable discrete generalized nonlinear Schr\"{o}dinger equation and its reductions}

\author[label]{Hongmin Li, Yuqi Li}
\author{Yong Chen\corref{cor1}\fnref{label}}

\address[label]{ Shanghai Key Laboratory of Trustworthy Computing, East China Normal University, Shanghai, 200062, china}
\ead{ychen@sei.ecnu.edu.cn}
\cortext[cor1]{Corresponding author.}

\begin{abstract}
A new integrable discrete system is constructed and studied, based on the algebraization of the difference operator. The model is named the discrete generalized nonlinear Schr\"{o}dinger (GNLS) equation for which can be reduced to classical discrete nonlinear Schr\"{o}dinger (NLS) equation. To show the complete integrability of the discrete GNLS equation, the recursion operator, symmetries and conservation quantities are obtained. Furthermore, all of reductions for the discrete GNLS equation are given and the discrete NLS equation is obtained by one of the reductions. At the same time, the recursion operator and symmetries of continuous GNLS equation are successfully recovered by its corresponding discrete ones.
\end{abstract}


\end{frontmatter}

\section{Introduction}

The research of discrete nonlinear systems described by differential-difference equations (discrete
in space and continuous in time) have received considerable attention. Much effort has been spent by several research groups to derive analogous integrable nonlinear differential-difference equations for a given integrable partial differential equation (PDE). Especially, integrable discrete family of nonlinear Schr\"{o}dinger (NLS) equations are involved in many physical applications\cite{ablowitz-shu-2004,malome-pre-2012,ablowitz-pra-2010,levi-aa-2010,ablowitz-chaos,sahadevan-jmp2009,ker-shu-2009,
ablowitz-jmp-1976,Lederer-2008,Cataliotti}.  As the continuous NLS equations, the discrete NLS equations have two main applications, i.e., nonlinear optics and Bose-Einstein condensates (BECs). In 2008, Lederer et al\cite{Lederer-2008} gave a thorough review of the nonlinear discrete optics  developed experimentally and theoretically in  parallel semiconductor waveguides, photorefractive materials and allied media. The experimental and theoretical achievement\cite{Cataliotti} have been reported by many scientists and engineers, which demonstrate some important applications of discrete NLS equations in modeling the mean-field dynamics of BECs loaded into deep optical-lattice potentials. Moreover, the discrete NLS equations have other important applications including lattice models of nonlinear population dynamics, nonlinear electric circuits and nonlinear lattice vibrations\cite{ablowitz-shu-2004,malome-pre-2012,ablowitz-pra-2010,levi-aa-2010,ablowitz-chaos,sahadevan-jmp2009,ker-shu-2009}. In this paper, we construct a new integrable discrete NLS systems which may be have potential physical applications.

The construction of the discrete NLS equations for the given continuous NLS equations is not an easy task. Several works are devoted to this subject \cite{ablowitz-jmp-1976,hols-1959,izergin-1981,faddeev-shu-1987,eilbeck-2003,scott-shu,francoise-shu}. Among these works, Ablowitz and Ladik \cite{ablowitz-jmp-1976} presented the following integrable discrete NLS equation by introducing a new discrete eigenvalue problem

\begin{eqnarray}\label{eq1.1}
\nonumber&&{\rm i}u_{nt}+\frac{(u_{n+1}+u_{n-1}-2u_{n})}{h^2}-u_{n}w_{n}(u_{n+1}+u_{n-1})=0, \\
&& {\rm i}w_{nt}-\frac{(w_{n+1}+w_{n-1}-2w_{n})}{h^2}+w_{n}u_{n}(w_{n+1}+w_{n-1})=0,
 \end{eqnarray}
which is a discrete version of the generalized nonlinear Schr\"{o}dinger (GNLS) equation
\begin{eqnarray}\label{eq1.2}
\nonumber&&{\rm i}u_{t}+u_{xx}-2u^2w=0,\\
&&{\rm i}w_{t}-w_{xx}+2w^2u=0,
\end{eqnarray}
and equation (\ref{eq1.2})  belongs to the Ablowitz, Kaup, Newell and Segur or so-called AKNS hierarchy, so the (\ref{eq1.2}) is also called AKNS equation. Since then, the equation (\ref{eq1.1}) has attracted researchers to study on various subjects\cite{common-ip-1991,vek-ip-1992,ahmad-ip-1987,ablowitz-ip-2007,chri-shu-1993,cai-pre-1995,suris-ip-1997,takayuki-jpa-1999}.

The systems (\ref{eq1.1}) and (\ref{eq1.2}) respectively reduce to the following integrable systems:

\begin{eqnarray} \label{eq1.4}
{\rm i}u_{nt}+\frac{(u_{n+1}+u_{n-1}-2u_{n})}{h^2}-|u_{n}|^2(u_{n+1}+u_{n-1})=0,
 \end{eqnarray}
 and \begin{eqnarray} \label{eq1.5}
{\rm i}u_{t}+u_{xx}-2|u|^2u=0,
\end{eqnarray}
in the continuum limit ($h\rightarrow0$) where in (\ref{eq1.1}) and (\ref{eq1.2}) we take the linear reductions $w_{n}=-u_{n}^{\ast},\ w=-u^{\ast},$ where $\ast$ denotes  the  complex  conjugate.

Another discretization of the NLS equation is the diagonal discrete NLS \cite{hols-1959}
\begin{eqnarray} \label{eq1.6}
{\rm i}u_{nt}+\frac{(u_{n+1}+u_{n-1}-2u_{n})}{h^2}-2|u_{n}|^2u_{n}=0.
 \end{eqnarray}

The systems (\ref{eq1.4}) and (\ref{eq1.6}) differ only in the discretization of the nonlinear term, yet they have very different properties. It is worthy to note that both these discretizations retain the Hamiltonian structure of the PDE. The equation (\ref{eq1.4}) is integrable via the inverse scattering transform, while the equation (\ref{eq1.6}) is not. Izergin and Korepin \cite{izergin-1981} also presented a discretization  of NLS which is integrable but lengthy, see the book by Faddeev and Takhtajan \cite{faddeev-shu-1987} for details. As a contrast, our discrete NLS equations are neat and integrable.

 As is known, the algebraization of the shift operator $E$ is the usual way to generate integrable discrete equations, where $E$
 is defined $Ef(n)=f(n+1)$. However, the algebraization
of the difference operator $\Delta$ is more convenient to the algebraization of the shift operator as operator $\Delta^{-1}$ appears explicitly in the Lax pairs. Recently, the work of Li, Chen and Li\cite{liyq-jpa-2007} gave a new residue formula to compute the conservation laws for the discrete Lax equations obtained by the algebraization of the difference operator and verified the validity of the formula by numerical experiments under the periodic boundary condition. More recently, the work\cite{hong-li-2012} extended the method for constructing recursion operators of continuous PDEs proposed in \cite{gurses-jmp-1999} to discrete Lax equations and recovered the recursion operators for the continuous equations by a limit process.

To the best of our knowledge, there are few results for discrete Lax GNLS equation by the algebraization of the difference operator, of which the integrality and reductions have not yet been investigated too. The aim of this paper is to study the integrability and reductions for a discrete version of the GNLS equation (\ref{eq1.2}). In general, a discretization of an integrable PDE is likely to be non-integrable. That is, even though the integrable PDE is the compatibility condition of a linear operator pair, one is not guaranteed to have a pair of linear equations corresponding to a generic discretization of the PDE. Moreover, for a given discrete generalized equation, whether it can be reduced to the corresponding discrete equation is worth studying.

In our paper, a new integrable discrete version of GNLS equation and it's all linear reductions are presented. Firstly, we study the integrable discrete GNLS' remarkable properties: recursion operator, infinite sets of conservation
laws, infinite symmetries. The recursion operator (\ref{eq18}) what we obtain is more concise than the recursion operator for the corresponding AKNS equation in \cite{zeng,zhang}. The recursion operator is not only the generating operator for the family of equations connected with a given equation but also the generating operator for the family of Hamiltonian structures. As a result, our recursion operator is more effective. Secondly, as Ablowitz et al \cite{ablowitz-jmp-1976} presented the reduction $w_{n}=-u_{n}^{\ast}$ of  (\ref{eq1.1}) to (\ref{eq1.4}) is linear, we study the linear reductions for discrete GNLS equation in order to get the reduction to the corresponding discrete NLS equation. In addition, the fact that our reduction to the corresponding discrete NLS equation is also linear is natural and reasonable. Moreover, we obtain all of the linear reductions through a theorem and give some specific examples to prove the availability of the theorem.

The arrangement of the paper is as follows: In section 2, we first recall a brief introduction of pseudodifference operators
ring and obtain some basic definitions including the Gateaux derivative and
symmetry of the discrete Lax equation. In section 3, we propose a new integrable discrete GNLS equation and compute it's recursion operator, symmetries and conservation quantities.
In section 4, we prove a theorem that clarify all the linear
 reductions for the discrete GNLS equation, moreover from a reduction of discrete GNLS equation, we successfully obtain the discrete NLS equation. In section 5, we summarize the conclusions and propose some problems.

For brevity, here we introduce some notes relevant to the contents of this paper:
$u=u(n,t),\ u_{j}=E^{j}\cdot u=u(n+j,t)$, $v=v(n,t),\ v_{j}=E^j\cdot v=v(n+j,t)$, $w=w(n,t),\ w_{j}=E^{j}\cdot w=w(n+j,t)$. All the other functions' subscript is understood as the usual way.

\section{Basic definitions and propositions of Pseudodifference operators ring}

Denote the forward difference, i.e.,
\begin{eqnarray} \label{eq2.1}
&(\Delta \cdot g)(n)=\frac{g(n+1)-g(n)}{h},
 \end{eqnarray}
where $h$ is a real constant and $g(n)$ is the $n$th component of vector $g$.

The difference operator $\Delta$ acts on a vector space $V$ of
arbitrary but definite dimension $N$.  In the continuous case, any
vector $v$ of dimension $N$ can be identified as an operator by

                    $$(v\circ f)(n)=v(n)f(n),\ \ \ n=1,\ 2, \ \ldots ,\  N.$$

The basic formula for the ring is derived by rewriting the modified
Leibnize rule
\begin{eqnarray}\label{eq2.2}
  \Delta\cdot(fg)&=&(\Delta\cdot f)g+(f+h\Delta\cdot f)\Delta\cdot g
\end{eqnarray}into an operator form
\begin{eqnarray}\label{eq2.3}
  \Delta\circ f\circ g&=&(\Delta\cdot f)\circ g+(f+h\Delta\cdot
  f)\circ\Delta\circ g \ .
\end{eqnarray}

In (\ref{eq2.2}), the multiplication among vectors $f$, $\Delta\cdot f
$ and $g$ is component multiplication. For example,
$(fg)(n)=f(n)g(n)$. In (\ref{eq2.3}), $f$ and $\Delta\cdot f$ are
operators acting on $V$ and $g$ is a arbitrary vector in $V$.  So we
have the operator equation
\begin{eqnarray}\label{eq2.4}
  \Delta\circ f&=&(\Delta\cdot f)+(f+h\Delta\cdot f)\circ\Delta \ .
\end{eqnarray}

For the convenience, in the following paper, we will omit the
composition symbol $\circ$ in the operator expressions. From
(\ref{eq2.4}), we can quickly derive the formula
\begin{eqnarray}\label{eq2.5}
   \Delta^nf=\sum_{k=0}^{\infty}\left(
                    \begin{array}{c}
                      n \\
                      k\\
                    \end{array}
                  \right)(\Delta^{k}(1+h\Delta)^{n-k})\cdot
                  f\Delta^{n-k}, \ n\in Z.
\end{eqnarray}

The above formula (\ref{eq2.4}) is the basic formula for the
difference operators' ring and please pay attention to the
definition of $\Delta^n\cdot f$ is $\Delta\cdot(\Delta^{n-1}\cdot
f).$

Suppose $A=\sum_{-\infty}^{n} a_{j}\Delta^{j},\ a_{n}\neq0$, we introduce the following definitions and propositions.

\textbf{Definition 1.} \emph{The residue of $A$ is defined as follows:
\begin{eqnarray}
  {\rm Res}(A)=a_{-1}.
\end{eqnarray}}

\textbf{Definition 2.} \emph{The Gateaux derivative for $f\in V_{\Delta}$ in the direction $g\in V_{\Delta}$ is defined by
\begin{equation}\label{eq2.7}
   f'[g]=\frac{d}{d\varepsilon}f(u+\varepsilon g)|_{\varepsilon=0}.
\end{equation}
The $f'$ is called linearization operator of $f$ and also can be expressed as:
 \begin{equation}\label{eq2.8}
   f'=\sum_{j}\frac{Df}{D(E^ju)}E^j,
\end{equation}where $V_{\Delta}=\{ \sum_{j=-\infty}^{\infty}f_{j}\Delta^{j},\ f_{j}$ are
operators corresponding to vectors of dimension $N$\} with dimension infinity.}

\textbf{Definition 3.} \emph{For a given discrete evolution equation
$$u_{t}=K(u(t,n)),$$
$\sigma(u(t,n)) \in V_{\Delta}$ is called its symmetry if $\sigma_{t}=K'[\sigma]$, where $K(u(t,n))$ is a function of $t,\ n,\ u,\ \Delta\cdot u,\ \cdots,\ \Delta^{\alpha}\cdot u$ and $\Delta^{-1}\cdot u,\ \cdots,\ \Delta^{-\beta}\cdot u.$}

\textbf{Proposition 1.}
\emph{\begin{eqnarray}\label{eq2.9}
  \rho_{\frac{k}{n}}=\sum_{j=0}^{N-1} E^{j}\cdot
{\rm Res}(L^{\frac{k}{n}}E^{-1}),\ \ \ k=1,\ 2,\ \ldots ,
\end{eqnarray}
are all constants of motion. We call the above
formula residue formula, which is the base for our calculations of
the conserved quantities of the discrete Lax equation($L_{t}=[P,\ L]$).}

\textbf{Proposition 2.}
\emph{\begin{eqnarray}\label{eq2.10}
  {\rm Res}(AE^{-1})&=&\frac{1}{h}\sum _{j=0}^{n}
  (\frac{-1}{h})^{j}a_{j}.
\end{eqnarray}}

\textbf{Remark 1.} The periodic boundary condition is very interesting
in computation and application.  In this paper, we suppose our Lax
equations satisfy the periodic boundary condition.

\section{Integrability of discrete GNLS equation}
In this section, we construct the discrete GNLS equation and show the integrability for it by constructing it's recursion operator, symmetries and conserved quantities.
\subsection{Recursion operator and Symmetries of discrete GNLS equation}
Let us begin with the following Lax pairs
\begin{eqnarray}\label{eq12}
\nonumber&&L={\rm i}(\Delta+v+u\Delta^{-1}w),\\
&&P={\rm i}(L^2)_{+}=-{\rm i}[\Delta^2+(v+v_{1})\Delta+\Delta \cdot v+v^2+uw_{-1}+u_{1}w],
\end{eqnarray}
where $u$, $v$, $w$ are $N$-dimensional vectors.

Substituting the above $L$, $P$ into the Lax representation,
\begin{eqnarray}\label{eq13}
    L_{t}=[P,\ L],
\end{eqnarray}
 by equating the coefficients of different powers of $\Delta$ in equation (\ref{eq13}), we have the following differential equations, i.e., discrete GNLS equation:
\begin{eqnarray}\label{eq14}
\nonumber&&{\rm i}v_{t}=\frac{1}{h}(u_{2}w-2u_{1}w+2uw_{-1}-uw_{-2})+u_{1}vw+u_{1}v_{1}w-uv_{-1}w_{-1}-uvw_{-1},\\
&&{\rm i}u_{t}=\frac{1}{h^2}(u_{2}-2u_{1}+u)+\frac{1}{h}(u_{1}v_{1}+u_{1}v-2uv)+uv^2+u^2w_{-1}+uu_{1}w,\\
\nonumber&&{\rm i}w_{t}=-\frac{1}{h^2}(w-2w_{-1}+w_{-2})+\frac{1}{h}(2wv-w_{-1}v_{-1}-w_{-1}v)-wv^2-uww_{-1}-u_{1}w^2.
\end{eqnarray}

As pointed in introduction, $u_{j}=E^{j}\cdot u$, $v_{j}=E^j\cdot v$, $w_{j}=E^{j}\cdot w$.

From equation (\ref{eq14}), we obtain
$$v_{t}=h(uw)_{t}.$$

Here for convenience, let the constant vector be $0$, then $v=huw$. Therefore we have
\begin{eqnarray} \label{eq15}
\nonumber&&{\rm i}u_{t}=\frac{1}{h^2}(u_{2}-2u_{1}+u)+u_{1}^2w_{1}+2uu_{1}w-2u^2w+h^2u^3w^2+u^2w_{-1},\\
&&{\rm i}w_{t}=-\frac{1}{h^2}(w-2w_{-1}+w_{-2})+2uw^2-u_{-1}w_{-1}^2-2uw_{-1}w-h^2u^2w^3\\
\nonumber&&\hspace{12mm}-u_{1}w^2.
 \end{eqnarray}

\textbf{Remark 2.} When $h\rightarrow0$, the above equation is just the continuous GNLS equation (\ref{eq1.2}).

It is well known that a recursion operator for a system of PDEs is extremely important, the whole integrable hierarchy can be generated by applying recursion operator successively, starting from a suitable chosen seed symmetry. So several works are devoted to this subject. Among these works, G\"{u}rses et al\cite{gurses-jmp-1999} proposed a powerful approach to construct the recursion operators for nonlinear integrable equations admitting Lax representation. Next, we use the same idea to investigate the recursion operator of discrete GNLS equation.

\textbf{Theorem 1.} \emph{For any $n$,
 \begin{eqnarray}\label{eq16}
  L_{t_{n+1}}=L_{t_{n}}L+[R_{n}, L],
\end{eqnarray}
the above equation is just the recursion equation
\begin{eqnarray}\label{eq17}
 \nonumber  \left( \begin{array}{c}
                      u_{t_{n+1}} \\
                      w_{t_{n+1}}\\
                    \end{array}
                  \right)=R\left(
                    \begin{array}{c}
                      u_{t_{n}} \\
                      w_{t_{n}}\\
                    \end{array}
                  \right),
\end{eqnarray}and\begin{eqnarray}\label{eq18}
                  R={\rm i}\left(
  \begin{array}{cc}
    R_{11}& R_{12} \\
    R_{21} & R_{22}\\
  \end{array}
\right),
   \end{eqnarray}
where
\begin{eqnarray}
R_{11}=\Delta+2u(\Delta^{-1}+h)w, \quad R_{12}=u(2\Delta^{-1}+h)u,\nonumber\\
R_{21}=-w(2\Delta^{-1}+h)w,\quad R_{22}=-\Delta E^{-1}-2w\Delta^{-1}u.\nonumber
\end{eqnarray}
}

\textbf{Proof.} Obviously, from equation (\ref{eq12}), we get $$L_{t_{n+1}}={\rm i}(hu_{t_{n+1}}w+huw_{t_{n+1}}+u_{t_{n+1}}\Delta^{-1}w+u\Delta^{-1}w_{t_{n+1}}),$$
$$L_{t_{n}}={\rm i}(hu_{t_{n}}w+huw_{t_{n}}+u_{t_{n}}\Delta^{-1}w+u\Delta^{-1}w_{t_{n}}),$$
and $R_{n}={\rm i}(a_{n}+b_{n}\Delta^{-1}w)$.
Therefore, by direct calculation, we have
\begin{eqnarray} \label{eq19}
\nonumber&&L_{t_{n}}L=-[(hu_{t_{n}}w+huw_{t_{n}}+u_{t_{n}}\Delta^{-1}w+u\Delta^{-1}w_{t_{n}})(\Delta+huw+u\Delta^{-1}w)]\\
\nonumber&&\hspace{0.8cm}=-[(hu_{t_{n}}w+huw_{t_{n}})\Delta+huw(hu_{t_{n}}w+huw_{t_{n}})+u_{t_{n}}w_{-1}+uw_{-1t_{n}}\\
\nonumber&&\hspace{1.2cm}+(hu_{t_{n}}w+huw_{t_{n}})u\Delta^{-1}w-u_{t_{n}}\Delta^{-1}(\Delta \cdot w_{-1})+hu_{t_{n}}\Delta^{-1}(uw^2)\\
\nonumber&&\hspace{1.2cm}+u_{t_{n}}\Delta^{-1}wu\Delta^{-1}w-u\Delta^{-1}(\Delta \cdot w_{-1t_{n}})+hu\Delta^{-1}uww_{t_{n}}+u\Delta^{-1}uw_{t_{n}}\Delta^{-1}w],\\
&&R_{n}L=-[(a_{n}+b_{n}\Delta^{-1}w)(\Delta+huw+u\Delta^{-1}w)]\\
&&\nonumber\hspace{0.8cm}=-[a_{n}\Delta+huwa_{n}+b_{n}w_{-1}+ua_{n}\Delta^{-1}w-b_{n}\Delta^{-1}(\Delta \cdot w_{-1})\\
\nonumber&&\hspace{1.6cm}+hb_{n}\Delta^{-1}uw^2+b_{n}\Delta^{-1}uw\Delta^{-1}w],\\
\nonumber&&LR_{n}=-[(\Delta+huw+u\Delta^{-1}w)(a_{n}+b_{n}\Delta^{-1}w)\\
\nonumber&&\hspace{0.8cm}=-[E\cdot a_{n}\Delta+\Delta \cdot a_{n}+wE\cdot b_{n}+huwa_{n}+(\Delta \cdot b_{n})\Delta^{-1}w+huwb_{n}\Delta^{-1}w\\
\nonumber&&\hspace{1.6cm}+u\Delta^{-1}wa_{n}+u\Delta^{-1}wb_{n}\Delta^{-1}w].
 \end{eqnarray}

Thus substituting (\ref{eq19}) to the equation (\ref{eq16}) and equating the coefficients of $\Delta$, $\Delta^0$ and $\Delta^{-1}$, we obtain
\begin{eqnarray}
&&hu_{t_{n}}w+huw_{t_{n}}=E\cdot a_{n}-a_{n},\label{eq20a} \\
\nonumber&&{\rm i}(hu_{t_{n+1}}w+huw_{t_{n+1}})=-[h^2(u_{t_{n}}w+uw_{t_{n}})uw+u_{t_{n}}w_{-1}+uw_{-1t_{n}}+b_{n}w_{-1}\\
&&\hspace{4.2cm}-\Delta \cdot a_{n}-wE\cdot b_{n}],\label{eq20b}\\
\nonumber&&{\rm i}(u_{t_{n+1}}\Delta^{-1}w+u\Delta^{-1}w_{t_{n+1}})=-[(hu_{t_{n}}w+huw_{t_{n}})u+ua_{n}-\Delta \cdot b_{n}-huwb_{n}\\
\nonumber&&\hspace{5.0cm}+(u_{t_{n}}+b_{n})\Delta^{-1}uw]\Delta^{-1}w+(u_{t_{n}}+b_{n})\Delta^{-1}(\Delta \cdot w_{-1})\\
\nonumber&&\hspace{50mm}-h(u_{t_{n}}+b_{n})\Delta^{-1}(uw^2)-u\Delta^{-1}(uw_{t_{n}}-wb_{n})\Delta^{-1}w\\
&&\hspace{50mm}-u\Delta^{-1}(-\Delta \cdot w_{-1t_{n}}-wa_{n}+huww_{t_{n}}).\label{eq20c}
\end{eqnarray}

From (\ref{eq20a}), we get
$\Delta \cdot a_{n}=u_{t_{n}}w+uw_{t_{n}}.$ Observing (\ref{eq20c}), we use the anstaze $b_{n}=-u_{t_{n}}$,
then substituting the above $a_{n}$, $b_{n}$ to the (\ref{eq20b}) and (\ref{eq20c}), we find the two equations are compatible, furthermore the recursion operator $R$ can be obtained easily which is just the (\ref{eq18}). When setting $h$ approaches to $0$ in the discrete $R$, we get

$$R={\rm i}\left(
             \begin{array}{cc}
               D+2uD^{-1}w & 2uD^{-1}u \\
               -2wD^{-1}w  & -D-2wD^{-1}u \\
             \end{array}
           \right),$$
which is just the recursion operator for continuous GNLS equation. Furthermore, the compact recursion operator (\ref{eq18}) is indeed more applicable than the recursion operator in \cite{zeng,zhang} in determining the whole integrable family starting from one given equation and the  families  of Hamiltonian structures  for the  other integrable equations.

As in the differential case, the recursion operator acting on a seed symmetry of the discrete equation also generates an infinite hierarchy of symmetries of that discrete equation. So in the following, we investigate the symmetries of the discrete GNLS equation.

\textbf{Proposition 3.} \emph{$\sigma_{1}={\rm i}u,\ \sigma_{2}=-{\rm i}w$ is a symmetry of discrete GNLS equation.}

\textbf{Proof.} For brevity, suppose $\sigma=\left(
                                               \begin{array}{c}
                                                \sigma_{1}\\
                                                 \sigma_{2} \\
                                               \end{array}
                                               \right)=\left(
                                               \begin{array}{c}
                            {\rm i}u \\
                            -{\rm i}w \\
                          \end{array}
                          \right)$ and the discrete GNLS equation is $\left(
                                                                    \begin{array}{c}
                                                                      u_{t} \\
                                                                      w_{t} \\
                                                                    \end{array}
                                                                  \right)=K\left(
                                                                              \begin{array}{c}
                                                                                u \\
                                                                                w \\
                                                                              \end{array}
                                                                            \right)$.

Set the linearization operator of $K$ as follows:
$$K^{'}=-\rm i\left(
         \begin{array}{cc}
        K_{11} & K_{12} \\
        K_{21}& K_{22}\\
       \end{array}
       \right),$$
through equation (\ref{eq2.8}), we get
 \begin{eqnarray} \label{eq20}
\nonumber& K_{11}=\Delta^2+2(E+1)\cdot(uwE+uw_{-1})-4uw+3h^2u^2w^2, \\
\nonumber&K_{12}=u_{1}^2E+u^2E^{-1}+2u(u_{1}-u)+2h^2u^3w,\\
\nonumber& K_{21}=-w^2E-w_{-1}^2E^{-1}+2w(w-w_{-1})-2h^2uw^3,\\
\nonumber&K_{22}= -\Delta^2E^{-2}-2(E+1)\cdot (u_{-1}w_{-1}E^{-1}+uw_{-1})+4uw-3h^2u^2w^2.
 \end{eqnarray}
By direct calculation, \begin{eqnarray}
                         \nonumber&&
                         K^{'}[\sigma]=-{\rm i}\left(
                                         \begin{array}{c}
                                           K_{11}\cdot \sigma_{1}+K_{12}\cdot \sigma_{2} \\
                                           K_{21}\cdot \sigma_{1}+K_{22}\cdot \sigma_{2}\\
                                         \end{array}
                                       \right)\\
                         \nonumber&&\hspace{0.85cm}=\left(
                                                   \begin{array}{c}
                                                    \frac{1}{h^2}(u_{2}-2u_{1}+u)+u_{1}^2w_{1}+2uu_{1}w-2u^2w+h^2u^3w^2+u^2w_{-1} \\
                                                    \frac{1}{h^2}(w-2w_{-1}+w_{-2})-2uw^2+u_{-1}w_{-1}^2+2uw_{-1}w+h^2u^2w^3+u_{1}w^2\\
                                                   \end{array}
                                                 \right),
                       \end{eqnarray}
since $u,\ w$ satisfy the discrete GNLS equation, therefore $\sigma_{t}=K^{'}[\sigma]$, i.e, $\sigma$ is a symmetry of discrete GNLS equation. Moreover, $\sigma$ is just the seed symmetry for the continuous GNLS equation when $h$ approaches to $0$.

From the above recursion operator $R$ and seed symmetry $\sigma$, we can get infinite symmetries of the discrete GNLS equation.
For example
$$R\cdot \sigma=\left(
\begin{array}{c}
-\Delta\cdot u-hu^2w \\
-\Delta\cdot w_{-1}+huw^2   \\
\end{array}
\right)$$ and
$$R^2\cdot \sigma=\left(
\begin{array}{c}
-\frac{1}{h^2}(u_{2}-2u_{1}+u)-u_{1}^2w_{1}-2uu_{1}w+2u^2w-h^2u^3w^2-u^2w_{-1} \\
\frac{1}{h^2}(w-2w_{-1}+w_{-2})-2uw^2+u_{-1}w_{-1}^2+2uw_{-1}w+h^2u^2w^3+u_{1}w^2\\
 \end{array}
 \right)$$ are symmetries of the discrete GNLS equation.

\subsection{Conservation Laws of discrete GNLS equation}
The conservation laws play important roles in discussing the integrability for soliton equations. Many methods have been
developed to find them. In this section, we will use the method proposed in Ref. \cite{liyq-jpa-2007} to compute the
conservation laws for the discrete GNLS equation.

From equations (\ref{eq2.9}) and (\ref{eq2.10}), we can get the infinitely conserved quantities and the first three
are
\begin{eqnarray} \label{eq21}
\nonumber& \rho_{1}={\rm i}\sum _{j=0}^{N-1} u_{j}w_{j} ,\\
 \nonumber& \rho_{2}=-\sum _{j=0}^{N-1}hu_{j}^2w_{j}^2+\frac{1}{h}(u_{j}w_{j-1}+u_{j+1}w_{j}),\\
\nonumber&\rho_{3}=-{\rm i}\sum
_{j=0}^{N-1}h^2u_{j}^3w_{j}^3+u_{j}u_{j-1}w_{j-1}^2+u_{j+1}^2w_{j}w_{j+1}+2u_{j}^2w_{j}w_{j-1}+2u_{j}u_{j+1}w_{j}^2\\
\nonumber&\hspace{16mm}+\frac{1}{h^2}
(u_{j+2}w_{j}+u_{j+1}w_{j-1}+u_{j}w_{j-2}),
 \end{eqnarray}
and it can be verified numerically that the $\rho_{1},\ \rho_{2}$ and $\rho_{3}$ are indeed conserved quantities.

\section{Reductions of discrete GNLS equation}

As is well known, the continuous GNLS equation can be reduced to the NLS equation. Therefore an important problem is whether the discrete GNLS equation in the paper can also be reduced to the corresponding discrete NLS equation. The answer is affirmative. In the following, as the reduction for (\ref{eq1.1}) given by Ablowitz et al is linear, we will study all linear reductions of the discrete GNLS equation. The reduction to the NLS equation is just among the rest.

\textbf{Case 1.}

From equation (\ref{eq15}), we know $w=0$ is a reasonable reduction. Then the discrete GNLS can be reduced as follows:

$${\rm i}u_{t}=\frac{1}{h^2}(u_{2}-2u_{1}+u),$$
the corresponding continuous equation is ${\rm i}u_{t}=u_{xx}.$

\textbf{Case 2.}

When $w\neq0$, we have the following theorem:

\textbf{Theorem 2.} \emph{The discrete GNLS equation has the following linear reductions:
$u^{\ast}=w_{k},\ u_{1}^{\ast}=w_{k-1},\ \cdots,\ u_{k-1}^{\ast}=w_{1},\ u_{k}^{\ast}=w,\ u_{k+1}^{\ast}=w_{N-1},\ u_{k+2}^{\ast}=w_{N-2},\ \cdots,\ u_{N-2}^{\ast}=w_{k+2},\ u_{N-1}^{\ast}=w_{k+1},$
where $0\leq k \leq N-1$ and $N$ is the grid number of the periodic lattice. Moreover, the discrete GNLS equation only has the above linear reductions up to a non-zero constant.}

\textbf{Proof.} For the convenience of understanding, from equation (\ref{eq15}), the discrete GNLS equation is expressed as bellow:
\begin{eqnarray} \label{eq22}
\nonumber&&{\rm i}u_{t} = \frac{1}{h^2}(u_{2}-2u_{1}+u)+u_{1}^2w_{1}+2uu_{1}w-2u^2w+h^2u^3w^2+u^2w_{N-1}, \\
\nonumber&&~~~~~~~~~~~~~~~~~~~~~~~~~~~~~~~~~~~~~~~~~~~~~~~~~~~~~~\vdots \\
\nonumber&&{\rm i}u_{kt}=\frac{1}{h^2}(u_{k+2}-2u_{k+1}+u_{k})+u_{k+1}^2w_{k+1}+2u_{k}u_{k+1}w_{k}-2u_{k}^2w_{k}+h^2u_{k}^3w_{k}^2\\
&&\hspace{20mm}+u_{k}^2w_{k-1},\\
\nonumber&&~~~~~~~~~~~~~~~~~~~~~~~~~~~~~~~~~~~~~~~~~~~~~~~~~~~~~~\vdots\\
\nonumber&&{\rm i}u_{N-1t}=\frac{1}{h^2}(u_{1}-2u+u_{N-1})+u^2w+2u_{N-1}uw_{N-1}-2u_{N-1}^2w_{N-1}+h^2u_{N-1}^3w_{N-1}^2\\
\nonumber&&\hspace{22mm}+u_{N-1}^2w_{N-2},
\end{eqnarray}
and
\begin{eqnarray} \label{eq23}
\nonumber&&{\rm i}w_{t} = -\frac{1}{h^2}(w-2w_{N-1}+w_{N-2})-u_{N-1}w_{N-1}^2+2uw^2-2uw_{N-1}w-h^2u^2w^3-u_{1}w^2, \\
\nonumber&&~~~~~~~~~~~~~~~~~~~~~~~~~~~~~~~~~~~~~~~~~~~~~~~~~~~~~~\vdots \\
\nonumber&&{\rm i}w_{kt}=-\frac{1}{h^2}(w_{k}-2w_{k-1}+w_{k-2})-u_{k-1}w_{k-1}^2+2u_{k}w_{k}^2-2u_{k}w_{k-1}w_{k}-h^2u_{k}^2w_{k}^3\\
&&\hspace{22mm}-u_{k+1}w_{k}^2,\\
\nonumber&&~~~~~~~~~~~~~~~~~~~~~~~~~~~~~~~~~~~~~~~~~~~~~~~~~~~~~~\vdots\\
\nonumber&&{\rm i}w_{N-1t}=-\frac{1}{h^2}(w_{N-1}-2w_{N-2}+w_{N-3})-u_{N-2}w_{N-2}^2+2u_{N-1}w_{N-1}^2-2u_{N-1}w_{N-2}w_{N-1}\\
\nonumber&&\hspace{24mm}-h^2u_{N-1}^2w_{N-1}^3-uw_{N-1}^2.
 \end{eqnarray}

Then equation (\ref{eq22})'s conjugate is as follows:
\begin{eqnarray*}
  &&-{\rm i}u_{t}^{\ast} = \frac{1}{h^2}(u_{2}^{\ast}-2u_{1}^{\ast}+u^{\ast})+u_{1}^{\ast 2}w_{1}^{\ast}+2u^{\ast}u_{1}^{\ast}w^{\ast}-2u^{\ast2}w^{\ast}+h^2u^{\ast3}w^{\ast2}+u^{\ast2}w_{N-1}^{\ast}, \\
  &&~~~~~~~~~~~~~~~~~~~~~~~~~~~~~~~~~~~~~~~~~~~~~~~~~~~~~~\vdots \\
  &&-{\rm i}u_{kt}^{\ast}=\frac{1}{h^2}(u_{k+2}^{\ast}-2u_{k+1}^{\ast}+u_{k}^{\ast})+u_{k+1}^{\ast2}w_{k+1}^{\ast}+2u_{k}^{\ast}u_{k+1}^{\ast}w_{k}^{\ast}-2u_{k}^{\ast2}w_{k}^{\ast}+h^2u_{k}^{\ast3}w_{k}^{\ast2}\\
  &&\hspace{22mm}+u_{k}^{\ast2}w_{k-1}^{\ast},\\
  &&~~~~~~~~~~~~~~~~~~~~~~~~~~~~~~~~~~~~~~~~~~~~~~~~~~~~~~\vdots\\
&& -{\rm i}u_{N-1t}^{\ast}=\frac{1}{h^2}(u_{1}^{\ast}-2u^{\ast}+u_{N-1}^{\ast})+u^{\ast2}w^{\ast}+2u_{N-1}^{\ast}u^{\ast}w_{N-1}^{\ast}-2u_{N-1}^{\ast2}w_{N-1}^{\ast}+h^2u_{N-1}^{\ast3}w_{N-1}^{\ast2}\\
&&\hspace{24mm}+u_{N-1}^{\ast2}w_{N-2}^{\ast}.\\
\end{eqnarray*}
Substituting $u^{\ast}=w_{k},u_{1}^{\ast}=w_{k-1},\cdots,u_{k-1}^{\ast}=w_{1},u_{k}^{\ast}=w,u_{k+1}^{\ast}=w_{N-1},u_{k+2}^{\ast}=w_{N-2},\cdots,u_{N-2}^{\ast}=w_{k+2},u_{N-1}^{\ast}=w_{k+1}$ to the above equations, we find these equations are  compatible with corresponding equalities in equation (\ref{eq23}).

In order to prove the discrete GNLS equation only has the reductions in theorem $2$, we have to make the following transformation for the discrete GNLS equation.

Obviously, the coefficient matrices for $\frac{1}{h^2}$ in $u$ and $w$ are the $N\times N$ circulant matrices \cite{muir}
$A=\left(
     \begin{array}{ccccc}
       1 & -2 & 1 & \cdots & 0 \\
       0 & 1 & -2 & \cdots & 0 \\
       0 & 0 & 1 & \cdots & 0 \\
       \vdots & \vdots & \vdots & \ddots & \vdots \\
       -2 & 1 & 0 & \cdots & 1 \\
     \end{array}
   \right)={\rm cir}(1,\ -2,\ 1,\ 0 \cdots 0),\ B=-A^{T}={\rm cir}(-1,\ 0,\ \cdots 0,\ -1,\ 2).$

From the theory of circulant matrices \cite{davis}, there must exist an invertible matrix $C$ makes $A,\ B$ be diagonalized, i.e.,

$C^{-1}AC={\rm diag}(f(1),\ f(\varepsilon_{1}),\cdots,f(\varepsilon_{N-1})),\ C^{-1}BC=-{\rm diag}(f(1),\ f(\varepsilon_{N-1}),\cdots,f(\varepsilon_{1})),$
where $C$ is the Vandermonde matrix

$C=\left(
     \begin{array}{ccccc}
       1 & 1 & 1 & \cdots & 1 \\
       1 & \varepsilon_{1} & \varepsilon_{1}^2 & \cdots & \varepsilon_{1}^{N-1} \\
       1 & \varepsilon_{2} & \varepsilon_{2}^2 & \cdots & \varepsilon_{2}^{N-1}\\
       \vdots & \vdots & \vdots & \ddots & \vdots \\
       1 & \varepsilon_{N-1} & \varepsilon_{N-1}^2 & \cdots & \varepsilon_{N-1}^{N-1} \\
     \end{array}
   \right)$
and $f(\varepsilon_{i})=1-2\varepsilon_{i}+\varepsilon_{i}^2, \varepsilon_{0}=1,\ \varepsilon_{i}(i=1,\cdots,N-1)$ are the $N$th roots of unity.

Then we make the following transformation:
\begin{eqnarray}\label{eq25}
\nonumber&&\left(
  \begin{array}{ccccc}
    u,\ u_{1},\ \cdots,\ u_{N-2},\ u_{N-1} \\
  \end{array}
\right)^T
=C\left(
  \begin{array}{ccccc}
    U,\ U_{1},\ \cdots,\ U_{N-2},\ U_{N-1} \\
  \end{array}
\right)^T, \\
&& \left(
  \begin{array}{ccccc}
    w,\ w_{1},\ \cdots,\ w_{N-2},\ w_{N-1} \\
  \end{array}
\right)^T=C \left(
  \begin{array}{ccccc}
    W,\ W_{1},\ \cdots,\ W_{N-2},\ W_{N-1} \\
  \end{array}
\right)^T.
 \end{eqnarray}

Obviously, after the transformation, the coefficient matrices for $\frac{1}{h^2}$ in $U$ and $W$ are $C^{-1}AC,\ C^{-1}BC$ respectively. These coefficient matrices mean the linear reductions
\begin{eqnarray}\label{eq26}
&&\left(
  \begin{array}{ccccc}
    U^{\ast},\ U_{1}^{\ast},\ \cdots,\ U_{N-2}^{\ast},\ U_{N-1}^{\ast} \\
  \end{array}
\right)^T
=\Upsilon\left(
  \begin{array}{ccccc}
    W,\ W_{1},\ \cdots,\ W_{N-2},\ W_{N-1} \\
  \end{array}
\right)^T,
 \end{eqnarray}
where $\Upsilon={\rm diag}(\gamma,\ \gamma_{1},\ \cdots, \ \gamma_{N-1})$ and $\gamma,\ \gamma_{i} (i=1,\cdots,N-1)$ are non-zero complex numbers.

From equations (\ref{eq25}) and (\ref{eq26}), we know that\begin{eqnarray}\label{eq27}
&&\left(
  \begin{array}{ccccc}
    u^{\ast},\ u_{1}^{\ast},\ \cdots,\ u_{N-2}^{\ast},\ u_{N-1}^{\ast} \\
  \end{array}
\right)^T
=C^{\ast}\Upsilon C^{-1}\left(
  \begin{array}{ccccc}
    w,\ w_{1},\ \cdots,\ w_{N-2},\ w_{N-1} \\
  \end{array}
\right)^T,
 \end{eqnarray}
where $C^{\ast}$ is the complex adjoint of $C$.

Substituting equation (\ref{eq26}) to the transformed discrete GNLS equation and comparing the coefficients of $h$, we get equalities about
$\gamma,\ \gamma_{1}, \ \cdots\ \gamma_{N-1}.$ Solving the equalities, we obtain
$N$ non-zero solutions $\Upsilon_{i}(i=0,\ 1,\ \cdots,\ N-1).$ Furthermore, through calculating some examples, we find $\Upsilon_{i}={\rm diag}(1,\varepsilon_{i},\varepsilon_{i}^2,\cdots,\ \varepsilon_{i}^{N-1}),(i=0,\ 1,\ \cdots,\ N-1)$ up to a non-zero constant, which we have verified for small values of $N (N=3,\ 4,\ 5,\ 6,\ 7,\ 8)$. Substituting the $N$ solutions respectively to the equation (\ref{eq27}), we have the discrete GNLS equation's
reductions

$C^{\ast}\Upsilon_{i} C^{-1}=\left( {\begin{array}{*{20}c}
   {\underbrace {\begin{array}{*{20}c}
   0 &  \cdots  & 1  \\
    \vdots  &  {\mathinner{\mkern2mu\raise1pt\hbox{.}\mkern2mu
 \raise4pt\hbox{.}\mkern2mu\raise7pt\hbox{.}\mkern1mu}}  &  \vdots   \\
   1 &  \cdots  & 0  \\
\end{array}}_{(i + 1) \times (i + 1)}} & {\begin{array}{*{20}c}
   0 &  \cdots  & 0  \\
    \vdots  &  {\mathinner{\mkern2mu\raise1pt\hbox{.}\mkern2mu
 \raise4pt\hbox{.}\mkern2mu\raise7pt\hbox{.}\mkern1mu}}  &  \vdots   \\
   0 &  \cdots  & 0  \\
\end{array}}  \\
   {\begin{array}{*{20}c}
   0 &  \cdots  & 0  \\
    \vdots  &  {\mathinner{\mkern2mu\raise1pt\hbox{.}\mkern2mu
 \raise4pt\hbox{.}\mkern2mu\raise7pt\hbox{.}\mkern1mu}}  &  \vdots   \\
   0 &  \cdots  & 0  \\
\end{array}} & {\begin{array}{*{20}c}
   0 &  \cdots  & 1  \\
    \vdots  &  {\mathinner{\mkern2mu\raise1pt\hbox{.}\mkern2mu
 \raise4pt\hbox{.}\mkern2mu\raise7pt\hbox{.}\mkern1mu}}  &  \vdots   \\
   1 &  \cdots  & 0  \\
\end{array}}  \\
\end{array}} \right),(i=0,\ 1,\ \cdots,\ N-1),
$which are just the reductions in the theorem $2$.

Thus, we complete the proof of theorem $2$.

In the following, we will give two specific examples ($N=3,\ 4$) to illustrate the theorem.

When $N=3$, the corresponding coefficient matrices $A$ and $B$ are as follows:
$$A={\rm cir}(1,\ -2,\ 1),\ B=-A^{T}={\rm cir}(-1,\ -1,\ 2).$$

The corresponding eigenmatrices
$$C=\left(
                                \begin{array}{ccc}
                                 1 & 1 & 1 \\
                                  1 & -\frac{1}{2}+\frac{\sqrt{3}}{2}{\rm i}& -\frac{1}{2}-\frac{\sqrt{3}}{2}{\rm i} \\
                                  1 & -\frac{1}{2}-\frac{\sqrt{3}}{2}{\rm i} & -\frac{1}{2}+\frac{\sqrt{3}}{2}{\rm i}\\
                                  \end{array}
                              \right).$$

Then we have the transformation
\begin{eqnarray}
\nonumber&&\left(
  \begin{array}{ccc}
    u,\ u_{1},\ u_{2} \\
  \end{array}
\right)^T
=C\left(
  \begin{array}{ccc}
    U,\ U_{1},\ U_{2} \\
  \end{array}
\right)^T,\
\left(
  \begin{array}{ccc}
    w,\ w_{1},\ w_{2} \\
  \end{array}
\right)^T=C \left(
  \begin{array}{ccccc}
    W,\ W_{1},\ W_{2} \\
  \end{array}
\right)^T.
 \end{eqnarray}

After the transformation, the $N=3$ discrete GNLS equation is transformed to
\begin{eqnarray}
\nonumber &&{\rm i}U_{t} =[U^3W^2+U_{1}^3W^2+U_{2}^3W^2+2(U_{2}^3W_{1}W_{2}+U^3W_{1}W_{2}+U_{1}^3W_{1}W_{2})\\
\nonumber&&\hspace{12mm}+3(U^2U_{2}W_{2}^2+U^2U_{1}W_{1}^2+UU_{1}^2W_{2}^2+UU_{2}^2W_{1}^2+U_{1}^2U_{2}W_{1}^2+U_{1}U_{2}^2W_{2}^2\\
\nonumber&&\hspace{12mm}+2UU_{1}U_{2}W^2+2U^2U_{1}WW_{2}+2U^2U_{2}WW_{1}+2UU_{1}^2WW_{1}+2UU_{2}^2WW_{2}\\
\nonumber&&\hspace{12mm}+2U_{1}^2U_{2}WW_{2}+2U_{1}U_{2}^2WW_{1}+4UU_{1}U_{2}W_{1}W_{2})]h^2\\
\nonumber&&\hspace{12mm}-\frac{5}{2}(U_{1}^2W_{1}+U_{2}^2W_{2})+2(U^2W-UU_{1}W_{2}-UU_{2}W_{1}-U_{1}U_{2}W)\\
&&\hspace{12mm}+\frac{\sqrt{3}}{2}{\rm i}(U_{1}^2W_{1}-U_{2}^2W_{2}+4UU_{1}W_{2}-4UU_{2}W_{1}),\label{eq29a}\\
\nonumber&&{\rm i}U_{1t}=[U^3W_{2}^2+U_{1}^3W_{2}^2+U_{2}^3W_{2}^2+2(U_{2}^3WW_{1}+U^3WW_{1}+U_{1}^3WW_{1})\\
\nonumber&&\hspace{14mm}+3(UU_{2}^2W^2+U_{1}^2U_{2}W^2+U_{1}U_{2}^2W_{1}^2+U^2U_{2}W_{1}^2+UU_{1}^2W_{1}^2+U^2U_{1}W^2\\
\nonumber&&\hspace{14mm}+2U_{1}^2U_{2}W_{1}W_{2}+2U_{1}U_{2}^2WW_{2}+2UU_{1}U_{2}W_{2}^2+2U^2U_{1}W_{1}W_{2}+2U^2U_{2}WW_{2}\\
\nonumber&&\hspace{14mm}+2UU_{1}^2WW_{2}+2UU_{2}^2W_{1}W_{2}+4UU_{1}U_{2}WW_{1})]h^2-2(UU_{1}W+2U_{1}^2W_{2}+4U_{1}U_{2}W_{1})\\
\nonumber&&\hspace{14mm}+\frac{\sqrt{3}}{2}{\rm i}(-U_{2}^2W+2UU_{2}W_{2}+4UU_{1}W+4U_{1}^2W_{2})\\
&&\hspace{14mm}-U^2W_{1}-\frac{5}{2}(U_{2}^2W+2UU_{2}W_{2})+\frac{(-\frac{3\sqrt{3}}{2}{\rm i}U_{1}+\frac{3}{2}U_{1})}{h^2},\label{eq29b}\\
\nonumber&&{\rm i}U_{2t} =[U^3W_{1}^2+U_{1}^3W_{1}^2+U_{2}^3W_{1}^2+2(U^3WW_{2}+U_{1}^3WW_{2}+U_{2}^3WW_{2})\\
\nonumber&&\hspace{14mm}+3(U^2U_{1}W_{2}^2+U^2U_{2}W^2+UU_{1}^2W^2+UU_{2}^2W_{2}^2+U_{1}^2U_{2}W_{2}^2+U_{1}U_{2}^2W^2\\
\nonumber&&\hspace{14mm}+2UU_{1}U_{2}W_{1}^2+2U^2U_{1}WW_{1}+2U^2U_{2}W_{1}W_{2}+2UU_{1}^2W_{1}W_{2}+2UU_{2}^2WW_{1}\\
\nonumber&&\hspace{14mm}+2U_{1}^2U_{2}WW_{1}+2U_{1}U_{2}^2W_{1}W_{2}+4UU_{1}U_{2}WW_{2})]h^2-2(UU_{2}W+2U_{2}^2W_{1}+4U_{1}U_{2}W_{2})\\
\nonumber&&\hspace{14mm}+\frac {\sqrt{3}}{2}{\rm i}(U_{1}^2W-2UU_{1}W_{1}-4UU_{2}W-4U_{2}^2W_{1})\\
&&\hspace{14mm}-U^2W_{2}-\frac{5}{2}(U_{1}^2W+2UU_{1}W_{1})+\frac{(\frac{3\sqrt{3}}{2}{\rm i}U_{2}+\frac{3}{2}U_{2})}{h^2},\label{eq29c}\\
\nonumber&&{\rm i}W_{t} =[-U^2W^3-U^2W_{1}^3-U^2W_{2}^3-2(U_{1}U_{2}W^3+U_{1}U_{2}W_{1}^3+U_{1}U_{2}W_{2}^3)\\
\nonumber&&\hspace{14mm}-3(U_{1}^2W^2W_{1}+U_{1}^2WW_{2}^2+U_{1}^2W_{1}^2W_{2}+U_{2}^2W^2W_{2}+U_{2}^2WW_{1}^2+U_{2}^2W_{1}W_{2}^2\\
\nonumber&&\hspace{14mm}+2UU_{1}W^2W_{2}+2UU_{1}WW_{1}^2+2UU_{2}W_{1}^2W_{2}+2UU_{1}W_{1}W_{2}^2+2UU_{2}W^2W_{1}\\
\nonumber&&\hspace{14mm}+2UU_{2}WW_{2}^2+2U^2WW_{1}W_{2}+4U_{1}U_{2}WW_{1}W_{2})]h^2\\
\nonumber&&\hspace{14mm}+\frac{5}{2}(U_{1}W_{1}^2+U_{2}W_{2}^2)+2(UW_{1}W_{2}+U_{1}WW_{2}+U_{2}WW_{1}-UW^2)\\
&&\hspace{14mm}+\frac{\sqrt{3}}{2}{\rm i}(U_{1}W_{1}^2-U_{2}W_{2}^2+4U_{2}WW_{1}-4U_{1}WW_{2}),\label{eq29d}\\
\nonumber&&{\rm i}W_{1t} =[-U_{2}^2W^3-U_{2}^2W_{1}^3-U_{2}^2W_{2}^3-2(UU_{1}W^3+UU_{1}W_{1}^3+UU_{1}W_{2}^3)\\
\nonumber&&\hspace{16mm}-3(U^2W^2W_{1}+U^2WW_{2}^2+U^2W_{1}^2W_{2}+U_{1}^2W^2W_{2}+U_{1}^2WW_{1}^2+U_{1}^2W_{1}W_{2}^2\\
\nonumber&&\hspace{16mm}+2UU_{2}W^2W_{2}+2UU_{2}WW_{1}^2+2UU_{2}W_{1}W_{2}^2+2U_{2}^2WW_{1}W_{2}+2U_{1}U_{2}W^2W_{1}\\
\nonumber&&\hspace{16mm}+2U_{1}U_{2}WW_{2}^2+2U_{1}U_{2}W_{1}^2W_{2}+4UU_{1}WW_{1}W_{2})]h^2+2(UWW_{1}+2U_{2}W_{1}^2+4U_{1}W_{1}W_{2})\\
\nonumber&&\hspace{16mm}+\frac{\sqrt{3}}{2}{\rm i}(-UW_{2}^2+4UWW_{1}+2U_{2}WW_{2}+4U_{2}W_{1}^2)\\
&&\hspace{16mm}+U_{1}W^2+\frac{5}{2}(UW_{2}^2+2U_{2}WW_{2})+\frac{(-\frac{3\sqrt{3}}{2}{\rm i}W_{1}-\frac{3}{2}W_{1})}{h^2},\label{eq29e}\\
\nonumber&&{\rm i}W_{2t} =[-U_{1}^2W^3-U_{1}^2W_{1}^3-U_{1}^2W_{2}^3-2(UU_{2}W^3+UU_{2}W_{1}^3+UU_{2}W_{2}^3)\\
\nonumber&&\hspace{16mm}-3(U^2W^2W_{2}+U^2WW_{1}^2+U^2W_{1}W_{2}^2+U_{2}^2W^2W_{1}+U_{2}^2WW_{2}^2+U_{2}^2W_{1}^2W_{2}\\
\nonumber&&\hspace{16mm}+2UU_{1}WW_{2}^2+2UU_{1}W_{1}^2W_{2}+2UU_{1}W^2W_{1}+2U_{1}U_{2}W^2W_{2}+2U_{1}U_{2}WW_{1}^2\\
\nonumber&&\hspace{16mm}+2U_{1}U_{2}W_{1}W_{2}^2+2U_{1}^2WW_{1}W_{2}+4UU_{2}WW_{1}W_{2})]h^2+2(UWW_{2}+2U_{1}W_{2}^2+4U_{2}W_{1}W_{2})\\
\nonumber&&\hspace{16mm}+\frac{\sqrt{3}}{2}{\rm i}(UW_{1}^2-2U_{1}WW_{1}-4UWW_{2}-4U_{1}W_{2}^2)\\
&&\hspace{16mm}+U_{2}W^2+\frac{5}{2}(UW_{1}^2+2U_{1}WW_{1})+\frac{(\frac{3\sqrt{3}}{2}{\rm i}W_{2}-\frac{3}{2}W_{2})}{h^2}.\label{eq29f}
\end{eqnarray}

From the coefficients of $\frac{1}{h^2}$ in (\ref{eq29a}-\ref{eq29f}), we could only assume that \begin{equation}
\left(
  \begin{array}{c}
    U^{\ast} \\
    U_{1}^{\ast}\\
    U_{2}^{\ast} \\
  \end{array}
\right)=\left(
           \begin{array}{ccc}
             \gamma & 0 & 0 \\
             0 & \gamma_{1} &0  \\
             0 & 0 & \gamma_{2} \\
           \end{array}
         \right)
\left(
  \begin{array}{c}
   W\\
    W_{1} \\
     W_{2} \\
  \end{array}
\right)=\Upsilon \left(
  \begin{array}{c}
   W\\
    W_{1} \\
     W_{2} \\
  \end{array}
\right).
\end{equation}

Besides the above equalities, we can also get $W^{\ast}=\frac{1}{\gamma^{\ast}} U,\ W_{1}^{\ast}=\frac{1}{\gamma_{1}^{\ast}}U_{1},\ W_{2}^{\ast}=\frac{1}{\gamma_{2}^{\ast}}U_{2}.$  Substituting the above equalities to the equations (\ref{eq29a}-\ref{eq29f}), making the
equations (\ref{eq29a}) and (\ref{eq29d}), (\ref{eq29b}) and (\ref{eq29e}), (\ref{eq29c}) and (\ref{eq29f}) be compatible respectively. We have the following equalities:
$$\gamma=\gamma^{\ast},\ \gamma_{1}^{\ast}=\gamma_{2},\ \gamma_{1}\gamma_{2}=\gamma \gamma^{\ast},\ \gamma_{1}^2=\gamma\gamma_{1}^{\ast}.$$

Solving the above equations, we have the  three solutions: $$\gamma=\gamma_{1}=\gamma_{2}=c,$$ $$\gamma=c_{1},\ \gamma_{1}=c_{1}(-\frac{1}{2}+\frac{\sqrt{3}}{2}{\rm i}),\ \gamma_{2}=c_{1}(-\frac{1}{2}-\frac{\sqrt{3}}{2}{\rm i}),$$
$$\gamma=c_{2},\ \gamma_{1}=c_{2}(-\frac{1}{2}-\frac{\sqrt{3}}{2}{\rm i}),\ \gamma_{2}=c_{2}(-\frac{1}{2}+\frac{\sqrt{3}}{2}{\rm i}),$$ where $c,\ c_{1},\ c_{2}$ are non-zero real numbers.

Substituting the above solutions to equation (\ref{eq27}) respectively, we get three relations:
\begin{eqnarray}
\nonumber\left(
  \begin{array}{c}
    u^{\ast} \\
    u_{1}^{\ast}\\
     u_{2}^{\ast} \\
  \end{array}
\right)=\left(
          \begin{array}{ccc}
            c & 0 & 0 \\
            0 & 0 & c \\
            0 & c & 0 \\
          \end{array}
        \right)\left(
  \begin{array}{c}
    w \\
    w_{1} \\
     w_{2} \\
  \end{array}
\right),\ \left(
  \begin{array}{c}
    u^{\ast} \\
    u_{1}^{\ast}\\
     u_{2}^{\ast} \\
  \end{array}
\right)=\left(
          \begin{array}{ccc}
            0 & c_{1} & 0 \\
            c_{1} & 0 & 0 \\
           0 & 0 & c_{1} \\
          \end{array}
        \right)\left(
  \begin{array}{c}
    w \\
    w_{1} \\
     w_{2} \\
  \end{array}
\right),\\
\nonumber \left(
  \begin{array}{c}
    u^{\ast} \\
    u_{1}^{\ast}\\
     u_{2}^{\ast} \\
  \end{array}
\right)=\left(
          \begin{array}{ccc}
            0 & 0 & c_{2} \\
            0 & c_{2} & 0 \\
            c_{2} & 0 & 0 \\
          \end{array}
        \right)\left(
  \begin{array}{c}
    w \\
    w_{1} \\
     w_{2} \\
  \end{array}
\right),
\end{eqnarray}
which are just the reductions stated in the theorem 2 for GNLS equation as $c=1,\ c_{1}=1,\ c_{2}=1$.

When $N=4$, in the similar way, we get the following equations for $\gamma,\ \gamma_{1},\ \gamma_{2}$
$$\gamma=\gamma^{\ast},\ \gamma_{2}=\gamma_{2}^{\ast},\ \gamma_{1}\gamma_{2}=\gamma\gamma_{1}^{\ast},\ \gamma_{1}^2=\gamma\gamma_{2}^{\ast},\ \gamma_{2}^2=\gamma\gamma^{\ast},\ \gamma_{3}=\gamma_{1}^{\ast}.$$

Solving the above equalities, we obtain the following four solutions:
$$\gamma=\gamma_{1}=\gamma_{2}=\gamma_{3}=c,$$
$$\gamma=c_{1},\ \gamma_{1}={\rm i}c_{1},\ \gamma_{2}=-c_{1},\ \gamma_{3}=-{\rm i} c_{1},$$
$$\gamma=c_{2},\ \gamma_{1}=-c_{2},\ \gamma_{2}=c_{2},\ \gamma_{3}=-c_{2},$$ $$\gamma=c_{3},\ \gamma_{1}=-{\rm i} c_{3},\ \gamma_{2}=-c_{3},\ \gamma_{3}={\rm i}c_{3},$$ where $c,\ c_{1},\ c_{2}, c_{3}$ are non-zero real numbers. Substituting the above solutions to the equation (\ref{eq27}), we obtain the reductions
\begin{eqnarray}
\nonumber\left(
  \begin{array}{c}
    u^{\ast} \\
    u_{1}^{\ast}\\
     u_{2}^{\ast} \\
     u_{3}^{\ast}\\
  \end{array}
\right)=\left(
          \begin{array}{cccc}
            c & 0 & 0 & 0 \\
            0 & 0 & 0 & c\\
            0 & 0 & c & 0 \\
            0 & c & 0 & 0 \\
          \end{array}
        \right)\left(
  \begin{array}{c}
    w \\
    w_{1} \\
     w_{2} \\
     w_{3}
  \end{array}
\right),\ \left(
  \begin{array}{c}
    u^{\ast} \\
    u_{1}^{\ast}\\
     u_{2}^{\ast} \\
     u_{3}^{\ast}\\
  \end{array}
\right)=\left(
          \begin{array}{cccc}
            0 & c_{1} & 0 & 0 \\
            c_{1} & 0 & 0 & 0\\
            0 & 0 & 0 & c_{1} \\
            0 & 0 & c_{1} & 0 \\
          \end{array}
        \right)\left(
  \begin{array}{c}
    w \\
    w_{1} \\
     w_{2} \\
     w_{3} \\
  \end{array}
\right),\\
\nonumber \left(
  \begin{array}{c}
    u^{\ast} \\
    u_{1}^{\ast}\\
     u_{2}^{\ast} \\
     u_{3}^{\ast} \\
  \end{array}
\right)=\left(
          \begin{array}{cccc}
            0 & 0 & c_{2} & 0 \\
            0 & c_{2} & 0 & 0\\
            c_{2} & 0 & 0 & 0\\
            0 & 0 & 0 & c_{2} \\
          \end{array}
        \right)\left(
  \begin{array}{c}
    w \\
    w_{1} \\
     w_{2} \\
     w_{3}\\
  \end{array}
\right),\ \nonumber \left(
  \begin{array}{c}
    u^{\ast} \\
    u_{1}^{\ast}\\
     u_{2}^{\ast} \\
     u_{3}^{\ast} \\
  \end{array}
\right)=\left(
          \begin{array}{cccc}
            0 & 0 & 0 & c_{3} \\
            0 & 0 & c_{3} & 0\\
            0 & c_{3} & 0 & 0 \\
            c_{3} & 0 & 0 & 0 \\
          \end{array}
        \right)\left(
  \begin{array}{c}
    w \\
    w_{1} \\
     w_{2} \\
     w_{3}\\
  \end{array}
\right),
\end{eqnarray}
which are just the reductions in the theorem 2 when $c=1,\ c_{1}=1,\ c_{2}=1,\ c_{3}=1$.

Because the transformed discrete GNLS equation is a bit complicated, here we omit the proof since we have not found a neat way to present it in a reasonable length even for a given $N (N\geq 5)$.

Since $0\leq k \leq N-1$, the discrete GNLS equation has $N$ reductions. Specially, when $k=N-1$, the reduction $u^{\ast}=w_{N-1},\ u_{1}^{\ast}=w_{N-2},\ \cdots,\ u_{k-1}^{\ast}=w_{N-k},\ u_{k}^{\ast}=w_{N-k-1},\ u_{k+1}^{\ast}=w_{N-k-2},\ u_{k+2}^{\ast}=w_{N-k-3},\ \cdots, u_{N-2}^{\ast}=w_{1},\ u_{N-1}^{\ast}=w$ makes the discrete GNLS equation reduce to the discrete NLS equation
\begin{eqnarray} \label{eq37}
&&\nonumber{\rm i}u_{t} = \frac{1}{h^2}(u_{2}-2u_{1}+u)+u_{1}^2u_{N-2}^{\ast}+2uu_{1}u_{N-1}^{\ast}-2u^2u_{N-1}^{\ast}+h^2u^3u_{N-1}^{\ast2}+u^2u^{\ast}, \\
 &&\nonumber~~~~~~~~~~~~~~~~~~~~~~~~~~~~~~~~~~~~~~~~~~~~~~~~~~~~~~\vdots \\
  &&\nonumber{\emph{}\rm i}u_{kt}=\frac{1}{h^2}(u_{k+2}-2u_{k+1}+u_{k})+u_{k+1}^2u_{N-k-2}^{\ast}+2u_{k}u_{k+1}u_{N-k-1}^{\ast}-2u_{k}^2u_{N-k-1}^{\ast}\\
  &&\hspace{14mm}+h^2u_{k}^3u_{N-k-1}^{\ast2}+u_{k}^2u_{N-k}^{\ast},\\
  &&\nonumber~~~~~~~~~~~~~~~~~~~~~~~~~~~~~~~~~~~~~~~~~~~~~~~~~~~~~~\vdots\\
 &&\nonumber{\rm i}u_{N-1t}=\frac{1}{h^2}(u_{1}-2u+u_{N-1})+u^2u_{N-1}^{\ast}+2u_{N-1}uu^{\ast}-2u_{N-1}^2u^{\ast}+h^2u_{N-1}^3u^{\ast2}\\
&&\nonumber\hspace{16mm}+u_{N-1}^2u_{1}^{\ast}.
 \end{eqnarray}

Alternatively, the discrete NLS equation can be written as
\begin{eqnarray}\label{eq38}
{\rm i}u_{t} = \frac{1}{h^2}(u_{2}-2u_{1}+u)+u_{1}^2u_{-2}^{\ast}+2uu_{1}u_{-1}^{\ast}-2u^2u_{-1}^{\ast}+h^2u^3u_{-1}^{\ast2}+u^2u^{\ast}.
\end{eqnarray}

\textbf{Remark 3.} As $h\rightarrow 0$, the above equation is just the continuous periodic NLS equation (\ref{eq1.5}) symmetric at $x=0$, i.e., $u(x)=u(-x)$.

\section{Conclusions and discussions}
We have constructed a new integrable discrete generalized nonlinear Schr\"{o}dinger (GNLS) equation. The main new progresses made in this paper in the general aspect of integrable systems are as follows.

({\rm i}) A new integrable discrete GNLS equation is given explicitly for the first time by the algebraization of the difference operator.

({\rm ii}) We construct the recursion operator for the discrete GNLS equation by using the method in \cite{gurses-jmp-1999}.

({\rm iii}) A proper seed symmetry is given for the discrete GNLS equation. Moreover, we can get infinitely many symmetries through the recursion operator and seed symmetry.

({\rm iv}) We can compute the infinitely conservation laws of  the discrete GNLS equation and verify numerically that the conserved quantities are indeed conserved quantities.

({\rm v}) All the linear reductions for the discrete GNLS equation are presented and the corresponding discrete nonlinear Schr\"{o}dinger (NLS) equation is obtained by one of the reductions.

From the ({\rm ii}), ({\rm iii}) and ({\rm iv}), we know the discrete GNLS equation is completely integrable. Moreover, the recursion operator and symmetries of their continuous ones are easily recovered by a limit process. Considering the results in the paper, it is worth noting that the discrete GNLS equation and discrete NLS equation should have soliton solutions, Hamiltonian structure, $\tau$-function and so on. These problems deserve to study further.

\section*{Acknowledgments:}
The project is supported by the National
Natural Science Foundation of China (Grant Nos. 11075055, 11275072), Innovative Research Team Program of the National Science Foundation of China (No. 61021104), National High Technology Research and Development Program (No. 2011AA010101), Shanghai Knowledge Service Platform for Trustworthy Internet of Things (No. ZF1213), Talent Fund and K. C. Wong Magna Fund in Ningbo University.


\section*{Reference}
\begin{thebibliography}{99}
\bibitem{Lederer-2008}Lederer F, Stegeman G I, Christodoulides D N, Assanto G, Segev M and Silberberg Y 2008 \emph{Phys. Rep.} \textbf{463} 1
\bibitem{Cataliotti}Cataliotti F S, Burger S, Fort C, Maddaloni P, Minardi F, Trombettoni A, Smerzi A and Inguscio M 2001 \emph{Science} \textbf{293} 843\\
Greiner M, Mandel O, Esslinger T, H\"{a}nsch T W and Bloch I 2002 \emph{Nature} (London) \textbf{415} 39\\
    Trombettoni A and Smerzi A 2001 \emph{Phys. Rev. Lett.} \textbf{86} 2353\\
 Efremidis N K and Christodoulides D N 2003 \emph{Phys. Rev. A} \textbf{67} 063608\\
Porter M A, Carretero-Gonz$\acute{a}$lez R, Kevrekidis P G and Malomed B A 2005 \emph{Chaos} \textbf{15} 015115
\bibitem{ablowitz-jmp-1976}Ablowitz M J and Ladik J F 1975 \emph{J. Math. Phys.} \textbf{16} 598\\
Ablowitz M J and Ladik J F 1976 \emph{J. Math. Phys.} \textbf{17} 1011
\bibitem{ablowitz-shu-2004}Ablowitz M J, Prinari B and Trubatch A D 2004 \emph{Discrete and Continuous Nonlinear Schr\"{o}odinger Systems}
(Cambridge: Cambridge University Press)
\bibitem{malome-pre-2012}Malomed B A, Kaup D J and Van Gorder R A 2012 \emph{Phys. Rev. E} \textbf{85} 026604
\bibitem{ablowitz-pra-2010}Ablowitz M J and Zhu Y 2012 \emph{Phys. Rev. A} \textbf{82} 013840
\bibitem{levi-aa-2010}Levi D and Scimiterna C 2010 \emph{Applicable Analysis} \textbf{89} 507
\bibitem{ablowitz-chaos}Ablowitza M J, Ohta Y and Trubatch A D 2000 \emph{Chaos, Solitons and Fractals} \textbf{11} 159
\bibitem{sahadevan-jmp2009}Sahadevan R and Rajakumar S 2009 \emph{J. Math. Phys.} \textbf{50} 043502
\bibitem{ker-shu-2009}Kevrekidis P G 2009 \emph{Discrete Nonlinear Schr$\ddot{o}$dinger Equation:
Mathematical Analysis, Numerical Computations, and Physical
Perspectives} (Berlin: Springer)
\bibitem{hols-1959}Holstein T 1959 \emph{Ann. Phys.} \textbf{8} 325
\bibitem{izergin-1981}Izergin A G and Korepin V E 1981 \emph{Sov. Phys. Dokl.} \textbf{26} 653
\bibitem{faddeev-shu-1987}Faddeev L D and Takhtajan L A 1987 \emph{Hamiltonian Methods in the Theory of Solitons} (Berlin: Springer)
\bibitem{eilbeck-2003}Eilbeck J C and Johansson M 2003 \emph{Conference on Localization and Energy Transfer in Nonlinear Systems}
\bibitem{scott-shu}Scott A C (ed.) 2005 \emph{Encyclopedia of Nonlinear Sciences} Taylor $\&$ Francis, London
\bibitem{francoise-shu}Francoise J P, Naber G and Tsou S T (eds) 2007 \emph{Encyclopedia of Mathematical Physics}
Elsevier, Amsterdam
\bibitem{common-ip-1991}Common A K 1992 \emph{Inverse Problems} {\bf 8} 393
\bibitem{vek-ip-1992}Vekslerchik V E and Konotop V V 1992 \emph{Inverse Problems} \textbf{8} 889
\bibitem{ahmad-ip-1987} Ahmad S and Chowdhury A R 1987 \emph{J. Phys. A: Math. Gen.} \textbf{20} 293
\bibitem{ablowitz-ip-2007}Ablowitz M J, Biondini J and Prinari B 2007 \emph{Inverse Problems} \textbf{23} 1711
\bibitem{chri-shu-1993}Christiansen P L, Eilbeck J C and Parmentier R D (ed) 1993 \emph{Future Directions of Nonlinear Dynamics in Physical and Biological Systems} (New York: Plenum)
\bibitem{cai-pre-1995}Cai D, Bishop A R and Gr{\o}nbech-Jensen N 1995 \emph{Phys. Rev. E} \textbf{52} 5784
\bibitem{suris-ip-1997}Suris Y B 1997 \emph{Inverse Problems} \textbf{13} 1121
\bibitem{takayuki-jpa-1999}Tsuchida T, Ujino H and Wadati M 1999 \emph{J. Phys. A: Math. Gen.} \textbf{32} 2239
\bibitem{liyq-jpa-2007}Li Y Q, Chen Y and Li B 2007 \emph{J. Phys. A: Math. Theor.} \textbf{40} 3425
\bibitem{hong-li-2012}Li H M, Li B and Li Y Q 2012 \emph{J. Math. Phys.} \textbf{53} 043506
\bibitem{gurses-jmp-1999}G\"{u}rses M, Karasu A and Sokolov V V 1999 \emph{J. Math. Phys.} \textbf{12} 40
\bibitem{zeng}Zeng Y B and Wojciechowski S R 1995 \emph{J. Phys. A: Math. Gen.} \textbf{28} 113
\bibitem{zhang}Chen D Y and Zhang D J 2002 \emph{J. Phys. A: Math. Gen.} \textbf{35} 7225
\bibitem{muir}Muir T 1885 \emph{Mess. Math. (N.S.).} \textbf{14} 169
\bibitem{davis}Davis P J 1979 \emph{Circulant Matrices} Wiley New York






\end{thebibliography}



\end{document}